\documentclass[a4paper,12pt]{article}
\begin{document}
\textwidth 165mm \textheight 240mm \topmargin 0mm
\leftmargin 20mm
\addtolength{\textheight}{-\headheight}
\addtolength{\textheight}{-\headsep}
\input{psbox}
% A useful Journal macro
\def\Journal#1#2#3#4{{#1} {\bf #2}, #3 (#4)}
% Some useful journal names
\def\NCA{\em Nuovo Cimento}
\def\NIM{\em Nucl. Instrum. Methods}
\def\NIMA{{\em Nucl. Instrum. Methods} A}
\def\NPA{{\em Nucl. Phys.} A}
\def\NPB{{\em Nucl. Phys.} B}
\def\PLB{{\em Phys. Lett.}  B}
\def\PRL{\em Phys. Rev. Lett.}
\def\PRC{{\em Phys. Rev.} C}
\def\PRD{{\em Phys. Rev.} D}
\def\ZPC{{\em Z. Phys.} C}
\setcounter{page}{0}
\thispagestyle{empty}
\mbox{ }\hfill{\normalsize UCT-TP 230/96}\\
\mbox{ }\hfill{\normalsize \today}\\
\begin{center}
{\Large \bf Hadronic Ratios and the\\[.5cm]
Number of Projectile Participants.}\\[1.cm]
{\bf   Jean Cleymans and Azwinndini  Muronga}\\[2.0em]
Department of Physics,
University of Cape Town, Rondebosch 7700, \\
South Africa  \\
\end{center}
\vskip 1cm
\begin{abstract}
We investigate the  dependence of hadronic ratios on the number
of projectile participants
using a thermal model 
incorporating  exact baryon number and strangeness conservation.
A  comparison  is  made  with  results from $Au-Au$
collisions obtained at the BNL-AGS.
\end{abstract}

Preliminary results on
the dependence of  hadronic ratios on the number
of  projectile  participants have recently been 
presented 
by the E866 collaboration~\cite{E866} for relativistic $Au-Au$ collisions
at the BNL-AGS.
These results give insight into the behaviour of the produced hadronic
system as a function  of the baryon number and
of the size of the interaction volume.

It is the purpose of the present paper to 
analyze these  results
using a  thermal resonance gas model
at a fixed  temperature and a fixed baryon density. Our 
treatment differs from previous~\cite{stachel,elliott} ones 
in that we consider the 
baryon content exactly.  This means that we do not 
introduce chemical potentials for
the baryon number (nor for strangeness). Chemical potentials are usually 
introduced to enforce the right quantum numbers of the system in
an average sense. This is a correct treatment for a large system, however,
for a small system the production of  e.g. an extra  
proton - anti-proton pair 
will clearly be more suppressed than in a large system. 
These extra corrections
were first pointed out by
Hagedorn~\cite{hagedorn}  and subsequently a  complete treatment was
presented by many people~\cite{hagedorn-redlich,greiner,gorenstein,CSW}.
We emphasize that these corrections 
do  not contain any information about the dynamics. They simply
follow from baryon number conservation. 
These corrections must be taken into account before
considering  more involved models. It is also worth emphasizing
that they
do not involve any new parameters.

As an example we analyze the 
(preliminary) data 
recently  presented  by  the  E866 collaboration~\cite{E866} at
BNL.

The exact treatment of quantum numbers in statistical mechanics 
is obtained by projecting
the partition function onto  the desired values of $B$ and $S$
\begin{equation}
Z_{B,S}={1\over 2\pi}\int_0^{2\pi}d\phi\ e^{-iB\phi}
\;  {1\over 2\pi}\int_0^{2\pi} d\psi  e^{-iS\psi}
Z(T,\lambda_B,\lambda_S)
\end{equation}
where the usual fugacity factors $\lambda_B$ and $\lambda_S$ have been replaced by :
\begin{equation}
\lambda_B = e^{i\phi}~~~~~~~~~~ \lambda_S = e^{i\psi}.
\end{equation}
We will use
\begin{equation}
B = 2N_{pp}
\end{equation}
where $N_{pp}$
 is the number of projectile participants with the factor 2
reflecting the symmetry of the $Au-Au$ collision system.
As the contributions always come pairwise for particle and anti-particle
the fugacity factors will give rise to the cosine of the angle.
In the further treatment it is useful to
group all particles appearing in the Particle  Data Booklet~\cite{PDB}
into four categories
depending on their quantum numbers (we leave out charm and bottom).
$Z_K$ is the sum (given below) of 
all mesons having strangeness $\pm 1$ ($K,\bar{K}, K^*,\dots $), similarly
$Z_N$ is the sum of all baryons and anti-baryons having zero strangeness,
$Z_Y$ is the sum of all hyperons and anti-hyperons while $Z_0$ is the sum of all
non-strange mesons,
and so on :
\begin{eqnarray}
Z_K &=& \sum_{j\in |S|=1, |B|=0} V\int {d^3p\over (2\pi)^3}e^{-E_j/T} , \nonumber\\
Z_N &=& \sum_{j\in |S|=0, |B|=1} V\int {d^3p\over (2\pi)^3}e^{-E_j/T} ,\nonumber\\
Z_Y &=& \sum_{j\in |S|=1, |B|=1} V\int {d^3p\over (2\pi)^3}e^{-E_j/T} ,\nonumber\\
Z_0 &=& \sum_{j\in |S|=0, |B|=0} V\int {d^3p\over (2\pi)^3}e^{-E_j/T} .
\end{eqnarray}

We do not include cascade particles as their 
contribution is unimportant for the energy range 
under consideration and their inclusion considerably
 complicates the formalism.
Each term will be multiplied by the cosine
 of an angle, either $\phi$ or $\psi$,
in the case where two angles are 
needed (e.g. for the hyperons) one introduces
a new one, $\alpha$, using
\begin{equation}
\delta (\phi -\psi -\alpha ) 
=\sum_{n=-\infty}^{\infty}e^{in(\phi -\psi -\alpha )}.
\end{equation}
Using the integral representation of the modified Bessel
functions
\begin{equation}
I_n(z) ={1\over \pi}\int_0^{\pi} e^{z\cos\theta}\cos{n\theta}\;
d\theta
\end{equation}
one can write the partition function  as
\begin{equation}
Z_{B,S}= Z_0\sum_{n=-\infty}^{\infty} I_n(2Z_Y)I_{n+B}(2Z_N)I_{n}(2Z_K)
\end{equation}
In order to discuss the particle abundances
it is useful to introduce the following quantities~\cite{CSW}
\begin{eqnarray}
R_K &=& \sum_{n=-\infty}^{\infty}
                        I_n(2Z_Y)I_{n+B}(2Z_N)I_{n+1}(2Z_K) ,\nonumber\\
R_N &=& \sum_{n=-\infty}^{\infty}
                        I_{n+1}(2Z_Y)I_{n+B-1}(2Z_N)I_{n}(2Z_K) ,\nonumber\\
R_Y &=& \sum_{n=-\infty}^{\infty}
                        I_{n+1}(2Z_Y)I_{n+B}(2Z_N)I_{n}(2Z_K) ,\nonumber\\
R_{\bar{K}} &=& \sum_{n=-\infty}^{\infty}
                        I_n(2Z_Y)I_{n+B}(2Z_N)I_{n-1}(2Z_K) ,\nonumber\\
R_{\bar{N}} &=& \sum_{n=-\infty}^{\infty}
                        I_{n+1}(2Z_Y)I_{n+B+1}(2Z_N)I_{n}(2Z_K) ,\nonumber\\
R_{\bar{Y}} &=& \sum_{n=-\infty}^{\infty}
                        I_{n-1}(2Z_Y)I_{n+B}(2Z_N)I_{n}(2Z_K) .
\end{eqnarray}

If a particle, $i$, has strangeness 1 and baryon number 0,
 it's density will be given by
\begin{equation}
n_{i} = \left[Z_0{R_K\over Z_{B,S}}\right]
\int {d^3p\over (2\pi)^3}e^{-E_i/T} .
\end{equation}
while  a particle with strangeness 0 and baryon number 1, 
will have a density  given by
\begin{equation}
n_i = \left[Z_0{R_N\over Z_{B,S}}\right]
\int {d^3p\over (2\pi)^3}e^{-E_i/T} 
\end{equation}
All other particle densities are obtained by using the appropriate $R$ factor
given in equation (8).
The factor in square brackets  replaces 
the fugacity  in the usual
grand canonical ensemble treatment~\cite{stachel,elliott}.
Having thus determined all particle 
densities, we consider
the behaviour at freeze-out time. In this case all the resonances 
in the gas are allowed to decay 
into lighter stable particles. 
This means that each particle density is multiplied with
its appropriate branching ratio (indicated by $Br$ below). 
The abundances of  particles in the final state are thus
determined by :
\begin{eqnarray}
n_{\pi^+} = \sum n_i Br(i\rightarrow \pi^+)  ,\nonumber \\
n_{K^+} = \sum n_i Br(i\rightarrow K^+)   ,\nonumber \\
n_{\pi^-} = \sum n_i Br(i\rightarrow \pi^-)  ,\nonumber \\
n_{K^-} = \sum n_i Br(i\rightarrow K^-)   ,\nonumber \\
n_{p} = \sum n_i Br(i\rightarrow p)   ,\nonumber \\
n_{\bar{p}} = \sum n_i Br(i\rightarrow \bar{p}) .
\end{eqnarray}
where each sum runs over all particles contained in the hadronic gas.

The comparison with 
experimental 
results is shown in figures 1 to 4.
To compare with earlier 
calculations~\cite{stachel,elliott}
we keep the temperature $T$ and the 
baryon density $B/V$ fixed. This corresponds to keeping the 
baryon chemical potential  fixed in the 
standard hadronic gas calculations using the grand canonical ensemble.

 In figure 1 we compare our results with recent data from the 
AGS~\cite{E866,stephans}. Figure 1 shows the $K^+/\pi^+$ ratio.
As one can see the results obtained from our calculation 
show a steep rise with  $N_{pp}$ before  leveling off. The
dependence on the baryon density is  minimal in this case. This result is
confirmed by calculations done in the grand canonical ensemble which also
show that this ratio is 
almost independent of the baryon density~\cite{elliott}. 
We note that the experimental data indicate a slower rise than 
the model calculation. 

In figures 2, 3 and 4 
 we  show  the  $K^-/\pi^+$  ,  the $K^+/K^-$ and the $p/\pi^+$
 ratios. In each 
case good agreement is obtained  with the results of the E866
collaboration~\cite{E866}.  The  relevant temperature is around
$T\approx 100$MeV, the baryon
density  is in the range $B/V\approx $ 0.02 - 0.05 fm$^{-3}$. In
the grand canonical ensemble this corresponds to a baryon 
chemical potential of $\mu_B\approx 540$ MeV.
\\[0.5cm]
{\bf   Acknowledgments}  We  gratefully  acknowledge  the financial
support of
the  University of Cape Town (URC) and of the Foundation for Research
develeopment(Pretoria).  We  acknowledge  the  help  of Duncan
Elliott for his  analysis of the data in the grand canonical ensemble.

\newpage
{\bf Figure Captions.}
\\
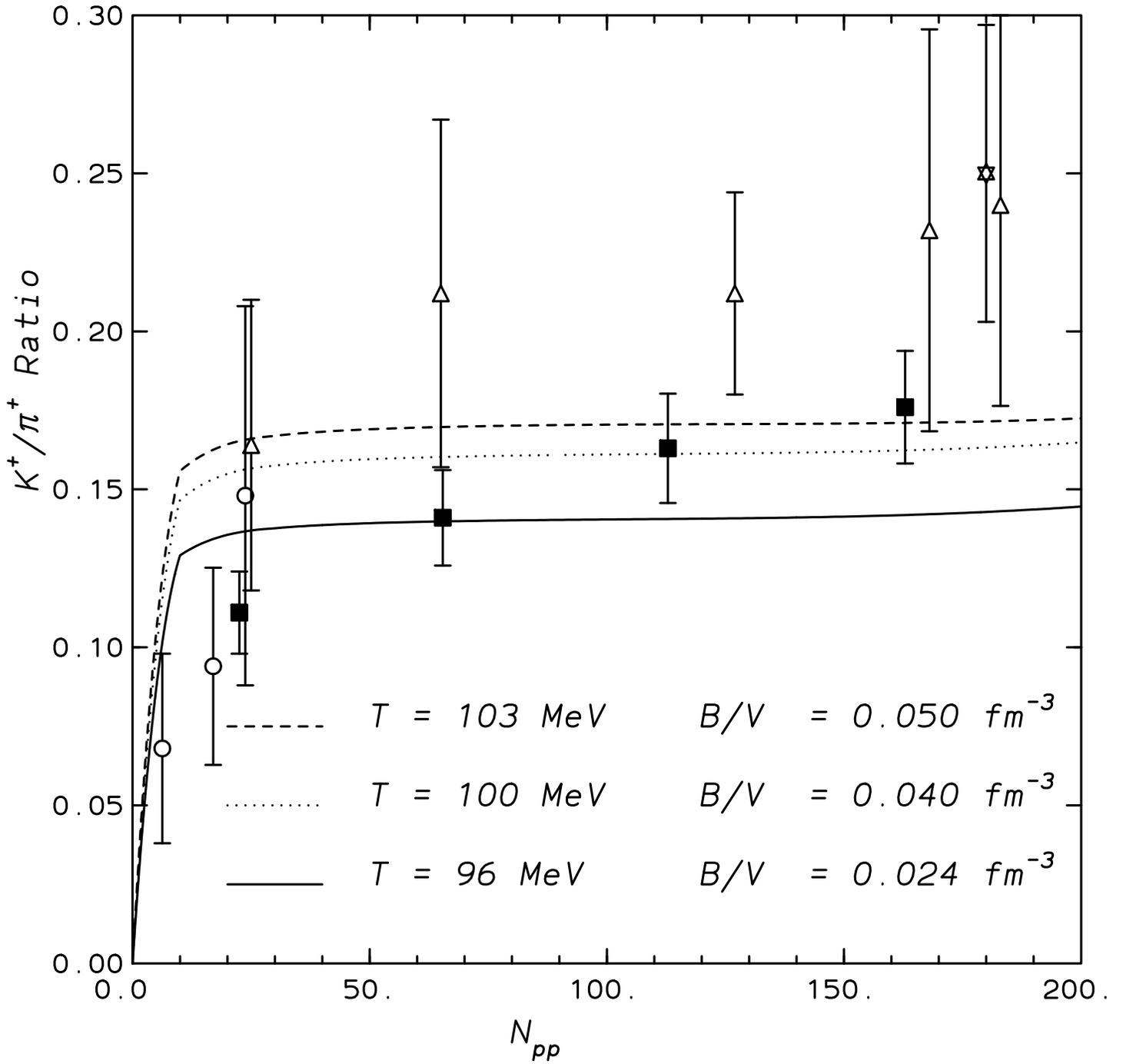
Figure 1 : The $K^+/\pi^+$
ratio as a function of the
 number of projectile participants $N_{pp}$. The solid line is 
obtained  for  $T$  =  96  MeV and $B/V$ = 0.024 fm$^{-3}$, the
dashed line corresponds to $T$ = 103 MeV and $B/V$ = 0.050 fm$^{-3}$
while  the dotted line corresponds to $T$ = 100 MeV and $B/V$ =
0.04 fm$^{-3}$.\\
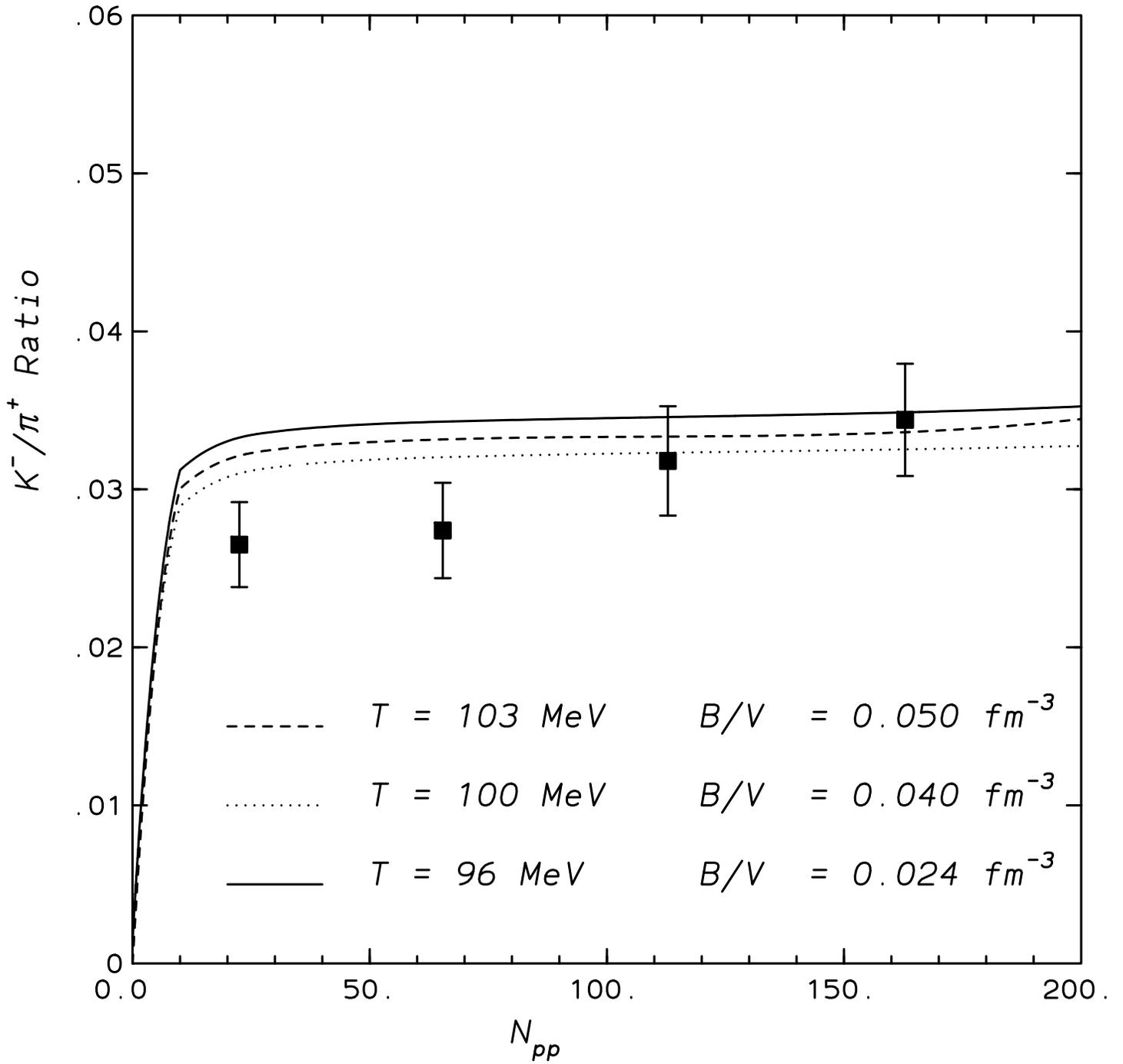
Figure 2 : The $K^-/\pi^+$
ratio as a function of the
 number of projectile participants $N_{pp}$.
The notation is the same as in figure 1. \\
Figure 3 : The $K^+/K^-$
ratio as a function of the
 number of projectile participants $N_{pp}$.
 The notation is the same as in figure 1.\\
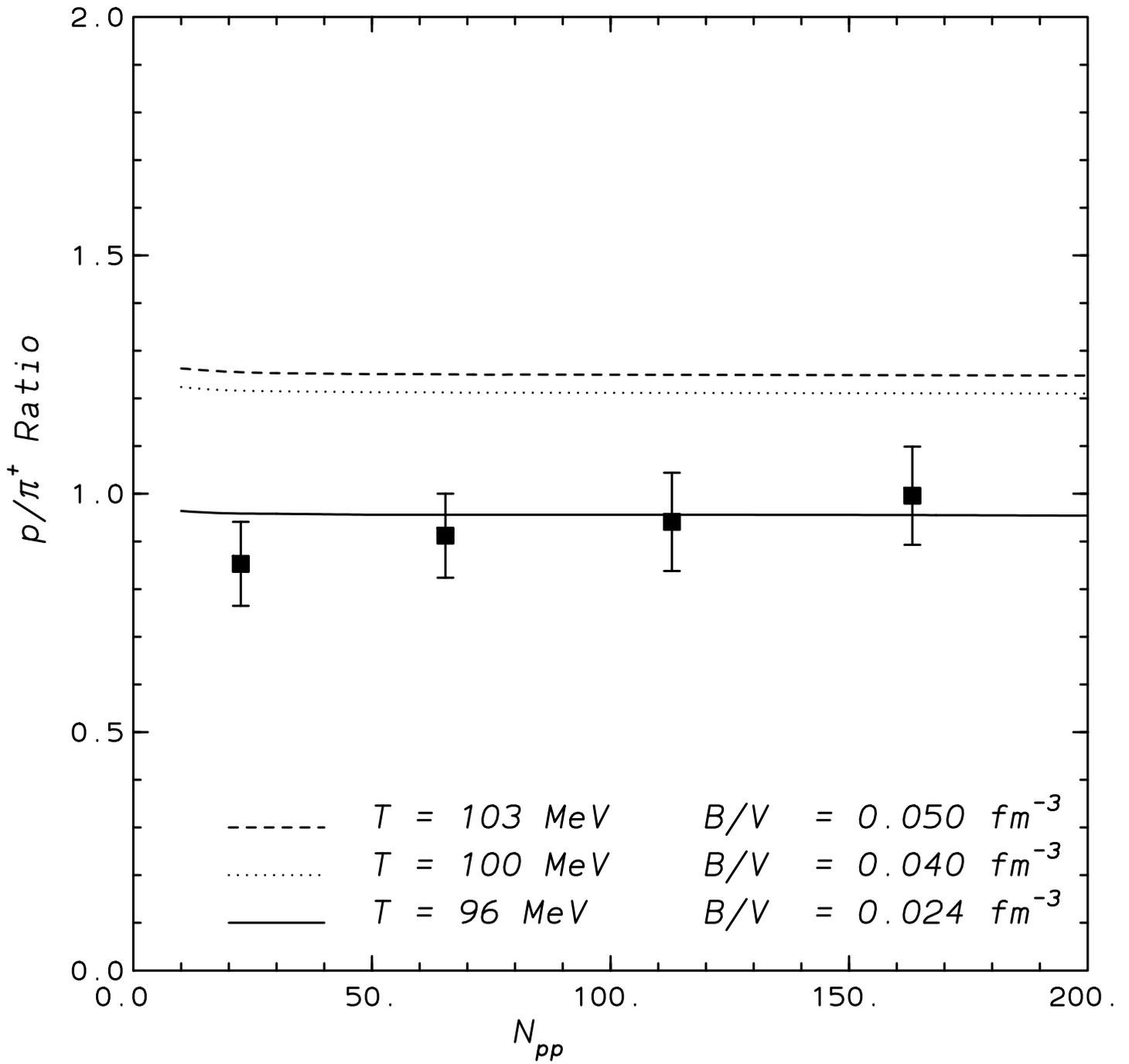
Figure 4 : The $p/\pi^+$
ratio as a function of the
 number of projectile participants $N_{pp}$.
 The notation is the same as in figure 1.\\
\end{document}